\documentclass[aps,prl,reprint,superscriptaddress,nobibnotes]{revtex4-2}

\usepackage{graphicx}
\usepackage{amsmath}
\usepackage{xcolor}

\begin{document}

\title{Collective dipole effects amplify field-driven ion transport\\
in amorphous Li$_3$PO$_4$}

\author{Haruka Doyu} \email[Contact author: ]{haruka.doyu@cello.t.u-tokyo.ac.jp}
\affiliation{Department of Applied Physics, Aalto University, Espoo, Finland}
\affiliation{Department of Materials Engineering, The University of Tokyo, Tokyo, Japan}

\author{Anh Khoa Augustin Lu}
\affiliation{Department of Materials Engineering, The University of Tokyo, Tokyo, Japan}
\affiliation{Research Center for Materials Nanoarchitectonics (MANA), National Institute for Materials Science (NIMS), Tsukuba, Japan}

\author{Satoshi Watanabe} \email[Contact author: ]{watanabe@cello.t.u-tokyo.ac.jp}
\affiliation{Department of Materials Engineering, The University of Tokyo, Tokyo, Japan}

\date{\today}

\begin{abstract}
Amorphous solid electrolytes often outperform their crystalline counterparts, but how disorder shapes ionic transport under strong electric fields is poorly understood. We use molecular dynamics with a machine-learned interatomic potential and a machine-learned Born-effective-charge model to measure Li mobility in amorphous Li$_3$PO$_4$ and in a crystal with Li vacancies, from the linear to the strongly nonlinear regime. The amorphous phase shows a much stronger field enhancement, and we trace this to a collective dipole effect. The migrating ion carries the same charge in both phases. Alone it would not make the amorphous phase respond more strongly. What makes the difference is the environment. In the amorphous network the surrounding atoms move with the hop. Their motion makes the work the field performs vary widely between hops. Since hopping rates grow exponentially with the work, the most strongly assisted hops can dominate the drift. Disorder thus enhances field-driven transport through the environment rather than the carrier itself, a mechanism that may operate in disordered ion conductors in general.
\end{abstract}

\maketitle

Amorphous solid electrolytes often conduct better than their crystalline counterparts~\cite{battery_overview,amorphous_oxyhalide}, but how structural disorder shapes ion transport under a strong electric field is poorly understood. A sufficiently strong field lowers the migration barriers themselves, and the mobility then grows nonlinearly with the field strength~\cite{cabrera,heuer2005,mobility_enhancement}. Such fields exceed those of battery operation but are reached in nanoscale ion-migration devices such as resistive-switching memories, where field-enhanced ion mobility is a recognized source of the nonlinearity these devices require~\cite{waser2009,vt_dilemma,mobility_enhancement,mobility_enhancement_recent}.

In many conventional descriptions of this regime, the field is assumed to perform work on the charge of the migrating ion alone. This assumption underlies the extensive simulations of crystalline conductors~\cite{mobility_enhancement,mobility_enhancement_recent,ambaum2026} as well as the treatments of disordered materials by hopping models on energy landscapes~\cite{heuer2014,wettstein2025}, by MD simulations of alkali silicate glasses~\cite{kunow2006}, and by experiments on ion-conducting glasses~\cite{heuer2005}. Within the modern theory of polarization, however, this work is set by the polarization change of the whole cell. The migrating ion and its polarized surroundings act as a collective dipole~\cite{salles}. Such collective dipole effects have been examined only for selected elementary processes in crystalline hosts~\cite{salles,abdallah2021,jia2022}. How they act in an amorphous network has remained unexplored, and this is the question studied here.

Lithium phosphate ($\text{Li}_3\text{PO}_4$) is a natural model system for this question. It is a well-established solid-state electrolyte whose amorphous phase attracts particular interest for its high ionic conductivity~\cite{li,exp_Ea}, and it serves as the solid electrolyte in lithium-ion batteries~\cite{battery_overview} and memory devices~\cite{memdev1,memdev2}. Computational studies have nevertheless focused on equilibrium conditions~\cite{li,du2008}, and simulations under external electric fields remain sparse, partly because they require both accurate interatomic forces and the atomic charges that couple the ions to the field~\cite{shimizu,nvt,hisama2023,sasaki2023,lu2026}. Notably, Shimizu \textit{et al.}~\cite{shimizu} demonstrated that amorphous $\text{Li}_3\text{PO}_4$ exhibits a higher mobility under an applied field than its crystalline counterpart with Li vacancies. That comparison, however, was made at a single high field ($E = 0.1$~V/\AA), where the mobility already contains a field-induced enhancement~\cite{mobility_enhancement}, and therefore mixes the intrinsic mobility with its field enhancement. The two must be separated across a broad range of field strengths, and the mechanism behind the amorphous advantage has remained elusive.

In this Letter, we separate the intrinsic mobilities from their field enhancement in amorphous $\text{Li}_3\text{PO}_4$ and a crystalline reference with Li vacancies, using molecular dynamics (MD) driven by a dual machine-learning framework. The amorphous phase shows both a larger intrinsic mobility and a stronger nonlinear field enhancement. We trace the stronger enhancement to the polarization work the field performs on each hop. The migrating ion carries the same charge in both phases and alone would not favor the amorphous phase. The stronger response comes instead from the surrounding network, which in the amorphous phase moves collectively with the hopping ion and makes the work vary strongly from hop to hop. This ability of the network to reorganize during a hop is not unique to $\text{Li}_3\text{PO}_4$, so the same mechanism may operate in other amorphous conductors.

The simulations rely on two machine-learning models. Zero-field interatomic forces are predicted by the GRACE interatomic potential~\cite{grace,gracespeed} (GRACE-2L-OAM variant), while the coupling to the field is described by per-atom Born effective charge (BEC) tensors predicted by the equivariant graph neural network Equivar~\cite{equivar}. The BEC tensor of atom $a$ describes the change of the macroscopic polarization $P_{\mu}$ with the atomic displacement $u_{a,\nu}$, $Z^{*}_{a,\mu\nu} = (\Omega/e)\, \partial P_{\mu}/\partial u_{a,\nu}$, where $\Omega$ is the cell volume. The total force on atom $a$ is $\mathbf{F}_{a} = \mathbf{F}^{0}_{a} + e\, Z^{*\mathsf{T}}_{a}\mathbf{E}$, where $\mathbf{F}^{0}_{a}$ is the zero-field interatomic force predicted by the potential and $\mathbf{E}$ is the applied electric field. The BEC tensors are evaluated at zero field. The coupling to the field is therefore linear, and higher-order electronic polarization is not included. The transport nonlinearity observed below arises instead from the exponential response of the hopping rates to the field-modified barriers. GRACE is fine-tuned on density functional theory (DFT) energies and forces, and Equivar on density functional perturbation theory (DFPT) BEC tensors, both computed with VASP~\cite{vasp1,vasp2} and originally generated in Refs.~\cite{li,shimizu}. The datasets, which cover pristine, Li-vacancy, and $\text{Li}_2\text{O}$-vacancy supercells sampled by \textit{ab initio} MD and nudged-elastic-band (NEB) calculations, are described in the Supplemental Material (SM)~\cite{SM}\nocite{PBE,ballistic,comp_Ea,he2017}. Both models were validated on held-out crystalline reference data, reaching root-mean-square errors of 0.84~meV/atom for energies, 16.7~meV/\AA{} for forces, and $0.032\,e$ for the BECs, and on an independent amorphous test set and against DFT-based NEB migration barriers (SM).

The production systems are an amorphous $\text{Li}_{384}\text{P}_{128}\text{O}_{512}$ cell (SM) and a crystalline $\text{Li}_{352}\text{P}_{128}\text{O}_{512}$ reference containing 8.3\% Li vacancies. The amorphous cell was generated by a melt-quench protocol with an equilibration temperature of 2500~K and a cooling rate of 0.01~K/fs at a density of $2.35~\text{g/cm}^3$, the DFT-relaxed density of Ref.~\cite{li}. The crystalline reference was constructed by removing 32 Li ions from a $\gamma$-$\text{Li}_3\text{PO}_4$ supercell~\cite{MatPro}, with the cartesian $x$, $y$, and $z$ axes along the crystallographic $a$, $b$, and $c$ directions, and the box volume fixed such that the density of the $\text{P}_{128}\text{O}_{512}$ framework is identical in both systems.

Canonical (NVT) MD simulations were performed with the Atomic Simulation Environment (ASE)~\cite{ase} using a 1~fs time step and a Nos\'e-Hoover thermostat~\cite{nosehoover} with a 100~fs coupling constant. Although the microcanonical ensemble is formally preferable under an external field, Joule heating restricts it to short runs and weak fields~\cite{nve}. Thermostatted nonequilibrium MD remains accurate provided the thermostat does not suppress the drift~\cite{nvt}. Following Ref.~\cite{nvt}, we verified that the measured drift is independent of the thermostat coupling constant over 10--200~fs (see SM).

\begin{figure}[t]
  \includegraphics[width=\columnwidth]{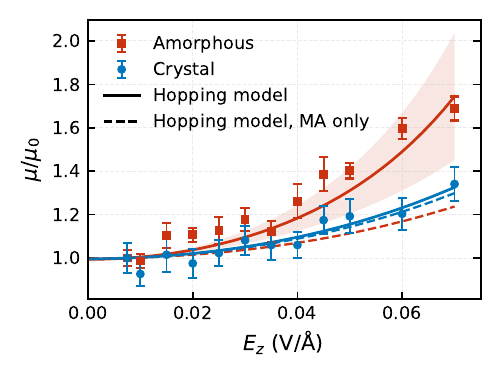}
  \caption{Normalized mobility $\mu/\mu_0$, with $\mu_0$ at $E_z = 0.0075$~V/\AA. Symbols are MD results, means of three runs with standard-error bars. Curves show the hopping model of the text, evaluated with the full polarization change (solid, with the bootstrap standard error as a band for the amorphous phase) and with the migrating-ion part alone (dashed).}
  \label{fig:linear}
\end{figure}

We measure Li transport at 900~K under applied electric fields, starting from equilibrated snapshots of the zero-field MD (SM). After 10~ps to reach a steady state under the applied field, each run is extended until the average Li displacement reaches at least 4~\AA. The field is applied along $z$, which for the crystal is the fastest diffusion direction of the anisotropic $\gamma$ phase, and Li displacements are measured relative to the center of mass of the $\text{P}_{128}\text{O}_{512}$ framework, which momentum conservation drives slowly opposite to the Li drift. From the steady-state drift velocity $v_d$ the mobility follows as $\mu = v_d/E$. The lowest field probed, $0.0075$~V/\AA, lies nearly in the linear regime of the measured field dependence [Fig.~\ref{fig:linear}]. We therefore take the mobility at this field as the low-field reference mobility $\mu_0$. Along $z$ we obtain $\mu_{0,\mathrm{am}} = (2.71 \pm 0.07) \times 10^{-4}$ and $\mu_{0,\mathrm{cry}} = (0.87 \pm 0.04) \times 10^{-4}$~cm$^2$/(V\,s). Averaging over the three field directions of the anisotropic crystal gives $(0.65 \pm 0.02) \times 10^{-4}$~cm$^2$/(V\,s) (SM). The amorphous phase is thus intrinsically four times more mobile than even a heavily defective crystal, consistent with experiment~\cite{kuwata,exp_Ea}, although the two cells also differ in carrier concentration. The superior transport reported at a single high field~\cite{shimizu} is therefore an intrinsic low-field property. The much stronger field response of the amorphous phase cannot be reduced to this advantage and is the central point that we address next.

Under fields up to $0.07$~V/\AA{} the normalized mobility $\mu/\mu_0$ grows exponentially in both phases, but much more steeply in the amorphous phase [Fig.~\ref{fig:linear}]. To explain this growth and its dependence on structure, we investigate elementary hops. Within transition-state theory, the applied field performs work on the migrating system. The work released between the initial state and the saddle point $S$ of migration path $i$ is given exactly by the change of the macroscopic polarization $\mathbf{P}$~\cite{kingsmith,resta,salles},
\begin{equation}
  W^{S}_{i} = \Omega\, \Delta\mathbf{P}^{S}_{i}\!\cdot\!\mathbf{E}.
  \label{eq:work}
\end{equation}
To first order in the field, i.e., along the zero-field path, the zero-field barrier $\Delta H_{i}$ of a symmetric two-way hopping channel is lowered to $\Delta H_{i} - W^{S}_{i}$ for hops along the field and raised to $\Delta H_{i} + W^{S}_{i}$ for hops against it. The net forward rate,
\begin{equation}
  \nu_{i}(E) \propto
  e^{-\Delta H_{i}/k_{B}T} \sinh\!\left(W^{S}_{i}/k_{B}T\right),
  \label{eq:rate}
\end{equation}
is linear in $E$ only while $W^{S}_{i} \ll k_{B}T$, so the mobility enhancement is controlled directly by how strongly the field modifies the barriers. An alternative source of nonlinearity in disordered conductors is a field-induced redistribution of carriers among inequivalent sites~\cite{heuer2014,wettstein2025}. The measured per-site Li occupancies, however, are unchanged within statistical error at every field in both phases (SM). Carrier redistribution can therefore be excluded as a significant contributor, supporting barrier modulation, i.e., the work of Eq.~\eqref{eq:work}, as the dominant origin of the enhancement. The question becomes which charge performs the work.

\begin{figure}[t]
  \includegraphics[width=\columnwidth]{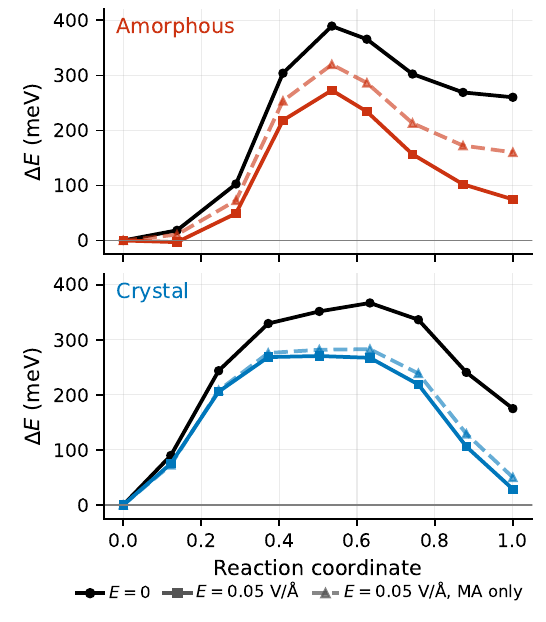}
  \caption{Example migration barriers in the two phases at $E = 0$ and
  under $E = 0.05$~V/\AA{} along each hop's forward direction, evaluated
  with the full polarization change (solid color) and with the migrating
  ion alone (dashed).}
  \label{fig:tilt}
\end{figure}

We suppress the path index $i$ below for brevity. The polarization change collects the response of every atom $a$ in the cell and separates exactly into the contribution of the migrating atom (MA) and that of its environment~\cite{salles},
\begin{equation}
\begin{split}
  \Delta\mathbf{P}^{S} &= \sum_{a} \Delta\mathbf{P}^{S}_{a}
  = \Delta\mathbf{P}^{S}_{\mathrm{MA}} + \Delta\mathbf{P}^{S}_{\mathrm{env}},\\
  \Delta\mathbf{P}^{S}_{a} &= \frac{e}{\Omega}\int_{0}^{S} Z^{*}_{a}\, d\mathbf{r}_{a},
\end{split}
  \label{eq:split}
\end{equation}
where the integral runs along the migration path, so that the work likewise splits, $W^{S} = W^{S}_{\mathrm{MA}} + W^{S}_{\mathrm{env}}$. To compare hops of different lengths on the scale of an ionic charge, we express each part as an equivalent charge obtained by projecting the polarization change along the migrating-ion displacement,
\begin{equation}
  Q_{\alpha} = \frac{\Omega\, \Delta\mathbf{P}^{S}_{\alpha}\!\cdot\!
  \mathbf{d}^{S}_{\mathrm{MA}}}{e\, |\mathbf{d}^{S}_{\mathrm{MA}}|^{2}},
  \qquad
  Q^{\mathrm{eqv}} = Q_{\mathrm{MA}} + Q_{\mathrm{env}},
  \label{eq:qeqv}
\end{equation}
where $\alpha = \mathrm{MA}, \mathrm{env}$ labels the two contributions and $\mathbf{d}^{S}_{\mathrm{MA}}$ is the displacement of the migrating atom from the initial state to the saddle point. $Q^{\mathrm{eqv}}$ is the equivalent polarization-work charge of Salles \textit{et al.}~\cite{salles}, here evaluated from the full path integral. $Q_{\mathrm{MA}}$ is essentially the projected BEC of the migrating ion, close to $+1$ for Li, and $Q_{\mathrm{env}}$ measures how much additional charge a field along the hop effectively transports because the environment moves with the hop~\cite{salles}. In high-symmetry environments the environment contributions largely cancel, whereas in low-symmetry networks $Q_{\mathrm{env}}$ can become comparable to $Q_{\mathrm{MA}}$~\cite{salles}. Here we apply the framework to the full hop ensemble of a disordered network.

We computed $Q^{\mathrm{eqv}}$ for 100 amorphous and 50 crystal hops. Candidate hops were detected in the zero-field MD at 900~K. In the amorphous phase, hops were drawn at random from each 0.5~\AA{} interval of MD hop length between 1.5 and 3.5~\AA{} until 25 were accepted, and in the crystal from all detected events until 50 were accepted. For each hop, the endpoint configurations were relaxed, and the minimum energy path was obtained with climbing-image NEB using GRACE. A hop was accepted only if its band is a jump of the detected Li between two stable sites, with the destination free of other Li, both barriers above $k_{B}T$, and no atom moving more than the detected Li. Of the attempted hops, 10\% were accepted in the amorphous phase and 77\% in the crystal (SM). Equivar BECs were evaluated on every image, and the polarization change to the saddle point was obtained by integrating $\sum_a Z^{*}_{a}\, d\mathbf{r}_a$ along the band. The path-integrated charges were validated directly against DFPT on independent amorphous NEB paths (SM).

Figure~\ref{fig:tilt} illustrates the outcome for an amorphous hop with a pronounced environment contribution and a typical crystal hop. The zero-field barriers are lowered by the work of Eq.~\eqref{eq:work} for a field of 0.05~V/\AA{} applied along each hop's forward direction, evaluated once with the full $\Delta\mathbf{P}^{S}$ and once with the migrating-ion contribution alone. In the amorphous example, the full polarization work lowers the barrier by 117~meV, of which 47~meV is supplied by the environment. In the crystal, the environment contributes only 12~meV of the 96~meV barrier reduction. The gap between the two field-modified curves is the collective dipole effect.

\begin{figure}[t]
  \includegraphics[width=\columnwidth]{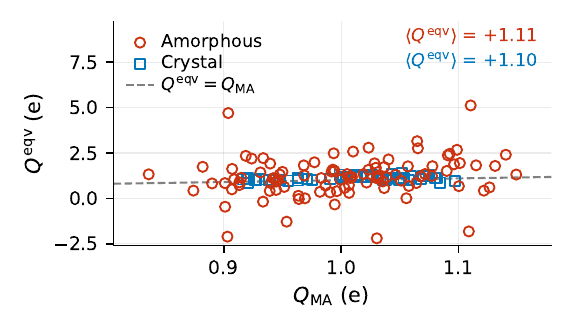}
  \caption{Equivalent charge $Q^{\mathrm{eqv}}$ against its migrating-ion
part $Q_{\mathrm{MA}}$ for the 100 amorphous and 50 crystal NEB hops. The dashed diagonal marks
  the migrating-ion-only line, $Q^{\mathrm{eqv}} = Q_{\mathrm{MA}}$, and the vertical
  distance from it is the environment contribution $Q_{\mathrm{env}}$.}
  \label{fig:qeqv}
\end{figure}

The full ensembles show how typical such a contribution is. Angle brackets denote the ensemble average taken as the ratio of sums, $\langle Q_{\alpha} \rangle = \Omega \sum_{i} \Delta\mathbf{P}^{S}_{\alpha,i}\!\cdot\! \hat{\mathbf{d}}^{S}_{\mathrm{MA},i} / (e \sum_{i} |\mathbf{d}^{S}_{\mathrm{MA},i}|)$ over the hops $i$, i.e., the total polarization work per unit total path length. This average splits exactly like Eq.~\eqref{eq:qeqv} and is robust against the few hops whose small saddle-point displacement inflates the per-hop ratio. The migrating ion itself is statistically identical in the two phases, with $\langle Q_{\mathrm{MA}}\rangle = +0.99 \pm 0.01$ in the amorphous phase and $+1.00 \pm 0.01$ in the crystal, where the uncertainties here and in the following are bootstrap standard errors over the hop ensembles. On average the environment contributes little in both phases. The mean equivalent charges agree within their uncertainties, $\langle Q^{\mathrm{eqv}}\rangle_{\mathrm{am}} = +1.11 \pm 0.10$ against $\langle Q^{\mathrm{eqv}}\rangle_{\mathrm{cry}} = +1.10 \pm 0.02$ [Fig.~\ref{fig:qeqv}], as do the environment terms, $\langle Q_{\mathrm{env}}\rangle = +0.11 \pm 0.10$ against $+0.09 \pm 0.01$. The two phases differ instead in their heterogeneity. Crystal hops hug the migrating-ion-only line with a hop-to-hop standard deviation of 0.11, whereas the amorphous values scatter with a standard deviation of 1.09, and 38\% of amorphous hops fall below the migrating-ion-only line, i.e., the environment there screens rather than amplifies.

The same ensemble connects the equivalent charges to the measured enhancement. Each hop is treated as an independent two-way channel whose forward and backward barriers are modified by the polarization work of Eq.~\eqref{eq:work}, and the resulting rates of Eq.~\eqref{eq:rate} are summed over the ensembles to form a hopping model for each phase (End Matter). Each hop enters with the displacement and polarization change of its NEB band, and in the isotropic amorphous phase it is averaged over all orientations, so neither model contains an adjustable parameter. Since the amorphous hops were drawn evenly over MD hop lengths, each amorphous hop is weighted by how often hops of its length occur in the zero-field MD (End Matter). As Fig.~\ref{fig:linear} shows, the model reproduces the measured enhancement of the crystal. In the amorphous phase the migrating ion alone gives $\mu/\mu_{0} = 1.24 \pm 0.03$ at 0.07~V/\AA, far below the measured 1.69, whereas the full polarization work gives $1.7 \pm 0.3$. The environment work thus accounts for the stronger amorphous response. The environment work raises the ensemble-averaged response in every hop-length range, most strongly for a minority of long hops with a large environment polarization (SM).

\begin{figure}[t]
  \includegraphics[width=\columnwidth]{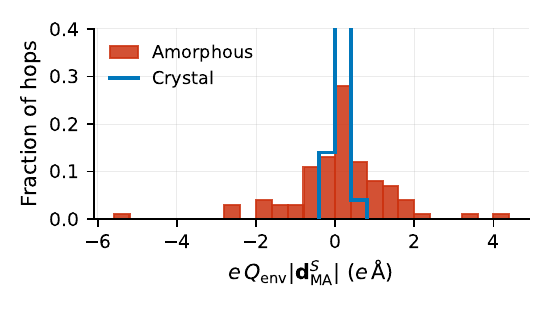}
  \caption{Distribution of the environment work per unit field along the hop, $e\,Q_{\mathrm{env}}|\mathbf{d}^{S}_{\mathrm{MA}}|$, over the 100 amorphous and 50 crystal NEB hops, as fractions of each ensemble. The crystal peak (0.82) is cut off.}
  \label{fig:workdist}
\end{figure}

Figure~\ref{fig:workdist} shows the distribution of the environment work per unit field along the hop, $e\,Q_{\mathrm{env}}|\mathbf{d}^{S}_{\mathrm{MA}}|$. Its mean is $+0.14~e\,$\AA{} in both phases, but its standard deviation is $1.23~e\,$\AA{} in the amorphous phase against $0.15~e\,$\AA{} in the crystal. Because the rates grow exponentially with the work, the hops in the strongly assisted tail dominate the response. Disorder thus acts through the width of this distribution rather than its mean. This comparison is a deliberately simple test of the collective dipole effect. It neglects, for example, the sampling bias, the hops excluded from the NEB ensemble, and the connectivity of the network (SM and End Matter).

The environment indeed moves far more in the amorphous phase (SM). Averaged over all hops, the summed displacement of the environment atoms at the saddle point is 26~\AA{} per hop against 8~\AA{} in the crystal. This motion can polarize the environment in two ways, by deforming or rotating the $\text{PO}_4$ units or by displacing the Li ions and the $\text{PO}_4$ units relative to one another. To separate the two, we recomputed the environment charge with each $\text{PO}_4$ unit forced to move as a whole, its five atoms sharing the displacement of their center of mass, which removes both the deformation and the rotation of each unit. The result reproduces $\langle Q_{\mathrm{env}}\rangle$ to within $0.02\,e$ in both phases. The difference between the recomputed and the actual charge has a standard deviation of only $0.1\,e$ in the amorphous phase and $0.04\,e$ in the crystal, against $1.1\,e$ and $0.1\,e$ for the actual charge. This displacement-based decomposition indicates that the collective dipole and its heterogeneity arise predominantly from relative motion of Li ions and $\text{PO}_4$ units rather than from deformation or rotation of the $\text{PO}_4$ units.

These results indicate that structural disorder amplifies field-driven transport through the environment rather than the charge carrier itself. The migrating-ion charge is fixed near $+1$ in both phases. What differs is how much of the surrounding network is set in motion by a hop, and above all how strongly this varies from hop to hop. Disorder creates a broad distribution of collective polarization work, and because the hopping rates depend exponentially on this work, its strongly assisted tail controls the nonlinear response. The same mechanism may operate broadly in disordered ion conductors when environmental reorganization generates a broad positive tail in the polarization work. Sulfide glasses such as amorphous $\text{Li}_3\text{PS}_4$ are natural systems in which to test it. The reorganization of the network is a design lever distinct from the usual carrier concentration and barrier height. Because the polarization work is obtained from zero-field NEB paths and Born effective charges alone, its statistics, the variance or the high-work tail, could serve as a descriptor for the strong-field response of candidate amorphous conductors. The fields at which the nonlinear regime sets in are reached in the nanometer-scale electrolyte layers of scaled ion-migration memories~\cite{memdev1,memdev2}, where the stronger amorphous response acts directly on the nonlinearity such devices require.

\begin{acknowledgments}
\textit{Acknowledgments.}---%
This work was supported by JSPS KAKENHI Grant No.~JP24K01284. The calculations were performed on the computational resources at the University of Tokyo and on the Wisteria/BDEC-01 supercomputer at the Information Technology Center, The University of Tokyo. The authors used Claude (Fable 5, Anthropic) to assist in drafting and revising the manuscript and in preparing plotting scripts. In addition, S.W. used ChatGPT (GPT-5.6 Sol, OpenAI) to critically review the manuscript and to obtain suggestions on its scientific logic, interpretation, presentation, and positioning relative to prior work. All scientific content, data, analyses, and conclusions are the authors' own. All AI-assisted output and suggestions were critically evaluated, independently verified where appropriate, and accepted or modified at the authors' discretion.
\end{acknowledgments}

\textit{Data availability}---The trained machine-learning models are openly available on Zenodo~\cite{zenodo_models}. All other data that support the findings of this study are available from the corresponding authors upon request.

\bibliography{references}

\onecolumngrid
\vspace{1em}
\begin{center}\textbf{\large End Matter}\end{center}
\vspace{0.5em}
\twocolumngrid
\setcounter{equation}{0}
\renewcommand{\theequation}{A\arabic{equation}}

\textit{Hopping model}---The model curves in Fig.~\ref{fig:linear} are obtained from the hop ensembles. Each hop is treated as an independent two-way channel whose forward and backward barriers change by $\mp W^{S}_{i}$, the polarization work of Eq.~\eqref{eq:work}. With the symmetric barrier of Eq.~\eqref{eq:rate}, the net rate is $\nu_{i} \propto e^{-\Delta H_{i}/k_{B}T} \sinh(W^{S}_{i}/k_{B}T)$. Each hop enters with the same barrier and attempt prefactor, so the equilibrium factor $e^{-\Delta H/k_{B}T}$ is a common prefactor absorbed into the proportionality. The drift velocity follows by summing the rates over the hops, weighted by their displacements along the field.

The crystal hops are drawn at random from all events of the zero-field MD, so equal weights follow the zero-field hopping rates. Since the crystal is anisotropic, each hop enters with its own geometry and with the field along $z$ as in the MD, $W^{S}_{i} = \Omega\,\Delta P^{S}_{z,i} E_{z}$, so that
\begin{equation}
  v(E_{z}) \propto \sum_{i} d_{z,i} \sinh\!\left(\frac{W^{S}_{i}}{k_{B}T}\right).
  \label{eq:ratesum}
\end{equation}
where $d_{z,i}$ is the $z$ component of $\mathbf{d}^{S}_{\mathrm{MA},i}$. For the isotropic amorphous phase, with 100 hops the high-field response would depend on which few strongly assisted hops happen to point along the field. Each hop is therefore rotated as a whole over all orientations, which are equally likely. The hops are drawn at random within each interval of MD hop length, and each hop is weighted by $w_{i}$, the fraction of the zero-field MD hops that fall in its interval (SM). For a field of strength $E$ in an arbitrary direction, the drift along the field is
\begin{equation}
  v(E) \propto \sum_{i} w_{i}\, d_{i}\cos\theta_{i}\, \sinh\!\left(\frac{\Omega\,|\Delta\mathbf{P}_{i}|\,E\cos\phi_{i}}{k_{B}T}\right),
  \label{eq:ratesum_dir}
\end{equation}
where $d_{i} = |\mathbf{d}^{S}_{\mathrm{MA},i}|$ is the displacement of hop $i$ to the saddle point, $\Delta\mathbf{P}_{i}$ its polarization change, $\theta_{i}$ the angle between the hop and the field, and $\phi_{i}$ the angle between $\Delta\mathbf{P}_{i}$ and the field. Averaged over all orientations, only the component of the hop direction along $\Delta\mathbf{P}_{i}$ survives (SM), which gives
\begin{equation}
  v(E) \propto \sum_{i} w_{i}\, d_{i}\cos\alpha_{i} \int_{-1}^{1} x \sinh(a_{i} x)\, dx,
  \label{eq:ratesum_iso}
\end{equation}
where $a_{i} = \Omega\,|\Delta\mathbf{P}_{i}|\,E/k_{B}T$, $\alpha_{i}$ is the angle between the hop and $\Delta\mathbf{P}_{i}$, and the integral equals $2(\cosh a_{i}/a_{i} - \sinh a_{i}/a_{i}^{2})$.

In both phases the curves show the mobility $\mu = v/E$ normalized at the lowest measured field $E_{0} = 0.0075$~V/\AA, i.e., $\mu(E)/\mu_{0}$ with $\mu_{0} = v(E_{0})/E_{0}$, exactly as for the measured data. Neither model contains an adjustable parameter. The band in Fig.~\ref{fig:linear} is the bootstrap standard error over the hops. Beyond this, the model assumes a fixed path geometry and neglects the connectivity of the network. Similar hopping models have nevertheless long been used to describe the nonlinear mobility of ionic conductors~\cite{mottgurney,heuer2005,mobility_enhancement}.

\end{document}

\title{Supplemental Material:\\ Collective dipole effects amplify field-driven
ion transport in amorphous Li$_3$PO$_4$}
\author{Haruka Doyu}
\affiliation{Department of Applied Physics, Aalto University, Espoo, Finland}
\affiliation{Department of Materials Engineering, The University of Tokyo, Tokyo, Japan}
\author{Anh Khoa Augustin Lu}
\affiliation{Department of Materials Engineering, The University of Tokyo, Tokyo, Japan}
\affiliation{Research Center for Materials Nanoarchitectonics (MANA), National Institute for Materials Science (NIMS), Tsukuba, Japan}
\author{Satoshi Watanabe}
\affiliation{Department of Materials Engineering, The University of Tokyo, Tokyo, Japan}
\date{\today}
\maketitle

\renewcommand{\thefigure}{S\arabic{figure}}
\renewcommand{\thetable}{S\arabic{table}}
\renewcommand{\theequation}{S\arabic{equation}}

\section{Models and training data}
\label{sec:sm-data}

Both models were obtained by fine-tuning published pretrained models. The interatomic potential is the two-layer GRACE-2L-OAM foundation model~\cite{grace,gracespeed} with a cutoff radius of 6~\AA{}. The BEC model is the base model~1 of Equivar~\cite{equivar}, with 6 interaction layers, a feature size of 64 in the interaction block, and a radial cutoff of 3~\AA{}.

Both models are trained on reference data originally generated in Refs.~\cite{li,shimizu}. The database comprises crystalline $\text{Li}_3\text{PO}_4$ supercells of varying size, covering pristine bulk, Li-vacancy, and $\text{Li}_2\text{O}$-vacancy configurations sampled by \textit{ab initio} MD (AIMD) and by NEB calculations. Reference energies and forces were computed with DFT and reference BEC tensors with DFPT, as implemented in VASP~\cite{vasp1,vasp2} with the PBE functional~\cite{PBE}. The plane-wave cutoff was 700~eV for the DFT calculations and 500~eV for the DFPT tensor evaluations. Tables~\ref{tab:mlip_data} and~\ref{tab:bec_data} summarize the dataset statistics.

For validation, the AIMD data were split sequentially. The first 90\% of each trajectory was used for training and the final 10\% was reserved for validation. The NEB data were split by complete pathways in the same 90/10 proportion.

\begin{table}[H]
\centering
\caption{Dataset statistics for the training of the interatomic potential.}
\label{tab:mlip_data}
\begin{tabular}{lllc}
\hline
\textbf{Structure} & \textbf{Defect} & \textbf{Type} & \textbf{Size} \\ \hline
$\text{Li}_6\text{P}_2\text{O}_8$ & No defects & AIMD (300, 2000, 4000 K) & 13\,000 \\
$\text{Li}_{12}\text{P}_4\text{O}_{16}$ & No defects & AIMD (300, 2000, 4000 K) & 13\,000 \\
$\text{Li}_{11}\text{P}_4\text{O}_{16}$ & Li vacancy & NEB & 1\,656 \\
$\text{Li}_{22}\text{P}_8\text{O}_{31}$ & $\text{Li}_2\text{O}$ vacancy & AIMD (2000 K) & 5\,000 \\
$\text{Li}_{46}\text{P}_{16}\text{O}_{63}$ & $\text{Li}_2\text{O}$ vacancy & AIMD (300 K) & 1\,873 \\ \hline
\end{tabular}
\end{table}

\begin{table}[H]
\centering
\caption{Dataset statistics for the training of the BEC model.}
\label{tab:bec_data}
\begin{tabular}{lllc}
\hline
\textbf{Structure} & \textbf{Defect} & \textbf{Type} & \textbf{Size} \\ \hline
$\text{Li}_{12}\text{P}_4\text{O}_{16}$ & No defects & AIMD (300, 2000 K) & 8\,000 \\
$\text{Li}_{11}\text{P}_4\text{O}_{16}$ & Li vacancy & NEB & 5\,000 \\
$\text{Li}_{22}\text{P}_8\text{O}_{31}$ & $\text{Li}_2\text{O}$ vacancy & AIMD (2000 K) & 4\,900 \\ \hline
\end{tabular}
\end{table}

\section{Model validation}
\label{sec:sm-validation}

\begin{figure}[htbp]
    \centering
    \includegraphics[width=0.45\linewidth]{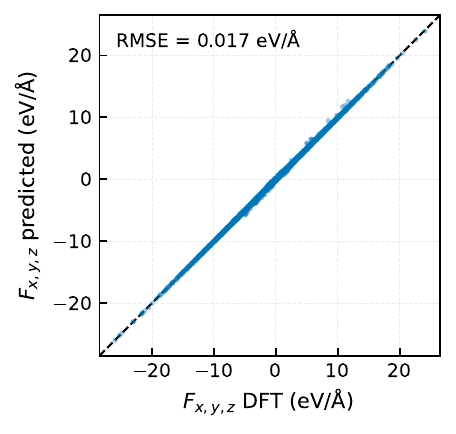}
    \caption{GRACE force components against the DFT reference on the crystalline validation set.}
    \label{fig:sm-force}
\end{figure}

\begin{figure}[htbp]
    \centering
    \includegraphics[width=0.95\linewidth]{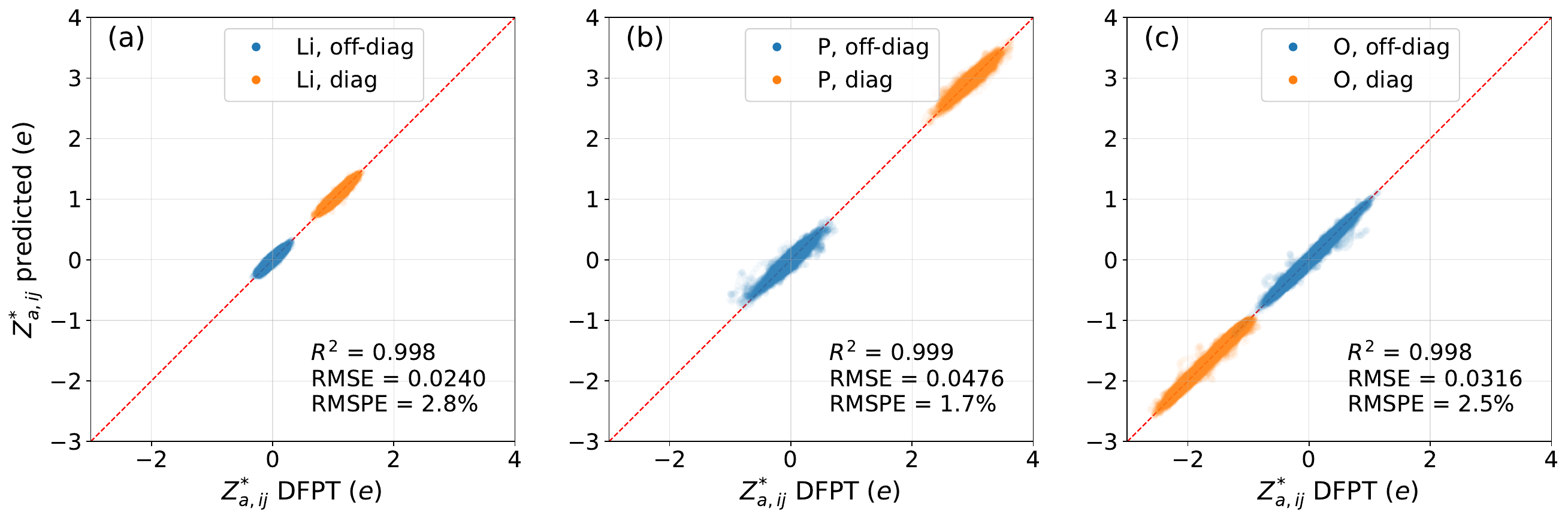}
    \caption{Equivar predictions of the BEC tensor components against the DFPT reference for (a)~Li, (b)~P, and (c)~O atoms in the crystalline validation set. The root-mean-square percentage error is evaluated on the diagonal elements of the charge tensors.}
    \label{fig:sm-bec}
\end{figure}

On the crystalline validation set GRACE reaches root-mean-square errors (RMSE) of 0.84~meV/atom for energies and 16.7~meV/\AA{} for forces, against 0.77 and 13.4 on the training set [Fig.~\ref{fig:sm-force}]. Equivar reaches a BEC RMSE of $0.032\,e$ ($0.025\,e$ on the training set) [Fig.~\ref{fig:sm-bec}]. For comparison, the earlier Behler--Parrinello networks of Ref.~\cite{shimizu} reached 2.9~meV/atom, 88~meV/\AA, and $0.038\,e$ on their test sets. The visible outliers for P and O in Fig.~\ref{fig:sm-bec} originate from $\text{P}_2\text{O}_7$ defects that occur only in the $\text{Li}_2\text{O}$-vacancy structures sampled at 2000~K, and such motifs are absent from the production simulations. No outliers appear in the force and energy predictions.

Since the training data are exclusively crystalline, the transferability of both models to the amorphous phase was tested on an independent structure. An initial $\gamma$-$\text{Li}_3\text{PO}_4$ configuration from the Materials Project~\cite{MatPro} was expanded into a 128-atom supercell at $\rho = 2.35~\text{g/cm}^3$, the DFT-relaxed density of Ref.~\cite{li}. One $\text{Li}_2\text{O}$ unit was removed because the pristine stoichiometry is difficult to maintain in a defect-free amorphous state at this cell size. The system was melted at 3000~K for 30~ps, quenched to 300~K at 1.0~K/fs, equilibrated for 5~ps, and relaxed until the maximum force fell below 0.01~eV/\AA. From an MD run of this structure at 900~K, 80 snapshots were extracted and recomputed with DFT and DFPT using a $4 \times 2 \times 2$ $k$-point mesh and the same settings as the training data. On this test set GRACE reaches 0.87~meV/atom for energies and 8.0~meV/\AA{} for forces, and Equivar reaches $0.027\,e$ with no discernible outliers, despite the presence of $\text{P}_2\text{O}_7$ motifs formed by the $\text{Li}_2\text{O}$ deficiency.

\begin{figure}[htbp]
    \centering
    \includegraphics[width=0.7\linewidth]{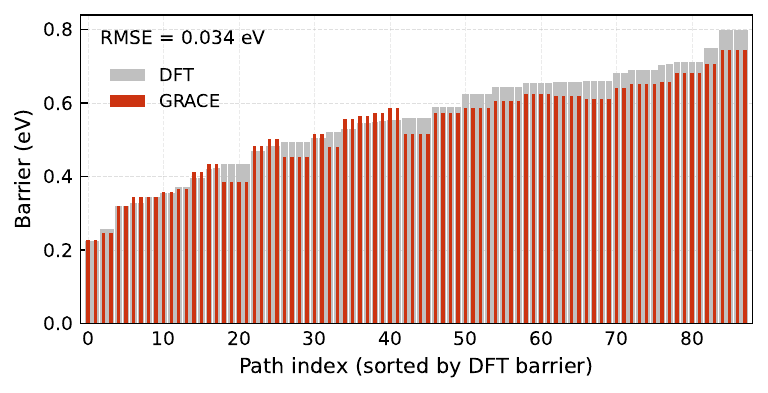}
    \caption{Migration barrier heights calculated with GRACE (red) and DFT (gray) for the 44 pathways evaluated in both forward and backward directions (88 barriers), sorted by the DFT barrier energy.}
    \label{fig:sm-neb}
\end{figure}

Migration barriers were validated separately with NEB calculations, following the barrier-validation scheme of Li \textit{et al.}~\cite{li} [Fig.~\ref{fig:sm-neb}]. A smaller 32-atom amorphous structure ($\rho = 2.35~\text{g/cm}^3$, melt-quenched from 4000~K without stoichiometric defects) was used. It partially crystallized due to the finite cell size but still provides a wide distribution of barrier heights. Diffusion paths were identified by selecting a target Li site and considering all Li ions within a 3.5~\AA{} radius as hopping candidates, which yields 44 unique paths. NEB calculations were performed with both GRACE and DFT (a $6 \times 6 \times 4$ $k$-point mesh) using seven intermediate images per path. GRACE reproduces the DFT barrier heights with an RMSE of 0.034~eV, with a small systematic underestimation of the largest barriers visible in Fig.~\ref{fig:sm-neb}.

\section{Production cells}
\label{sec:sm-cells}

\begin{figure}[htbp]
    \centering
    \includegraphics[width=0.85\linewidth]{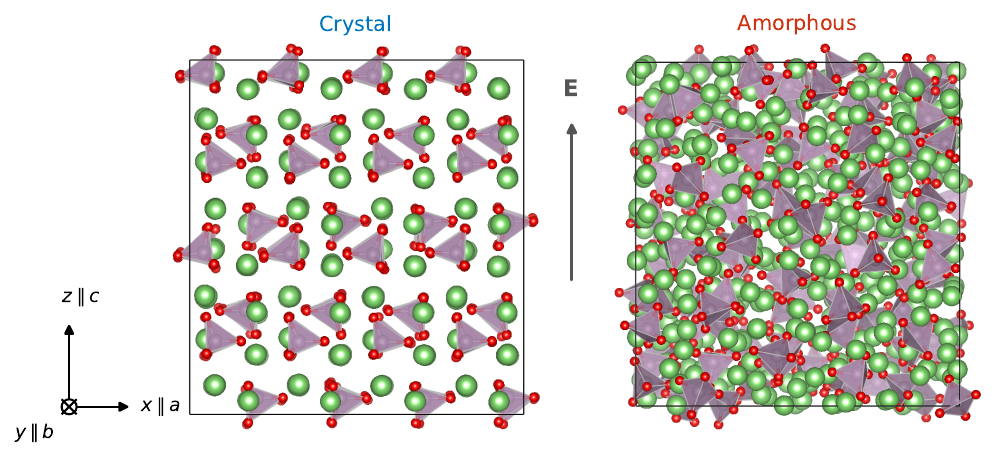}
    \caption{The relaxed production cells viewed along the $y$ axis. Left, the Li-vacancy crystal $\text{Li}_{352}\text{P}_{128}\text{O}_{512}$, with the Cartesian axes, their crystallographic directions, and the field direction indicated. Right, the amorphous $\text{Li}_{384}\text{P}_{128}\text{O}_{512}$ cell. Li ions are shown as green spheres and the $\text{PO}_4$ units as tetrahedra with O in red.}
    \label{fig:sm-structure}
\end{figure}

Figure~\ref{fig:sm-structure} shows the two production cells viewed along the $y$ axis. The left panel shows the Li-vacancy crystal ($\text{Li}_{352}\text{P}_{128}\text{O}_{512}$) together with the orientation of the Cartesian axes with respect to the crystallographic directions of the conventional cell~\cite{MatPro} and the direction of the applied field. The right panel shows the amorphous cell ($\text{Li}_{384}\text{P}_{128}\text{O}_{512}$) generated by the melt-quench protocol described in the main text.

\section{Thermostat validation}
\label{sec:sm-thermostat}

\begin{figure}[htbp]
    \centering
    \includegraphics[width=0.75\linewidth]{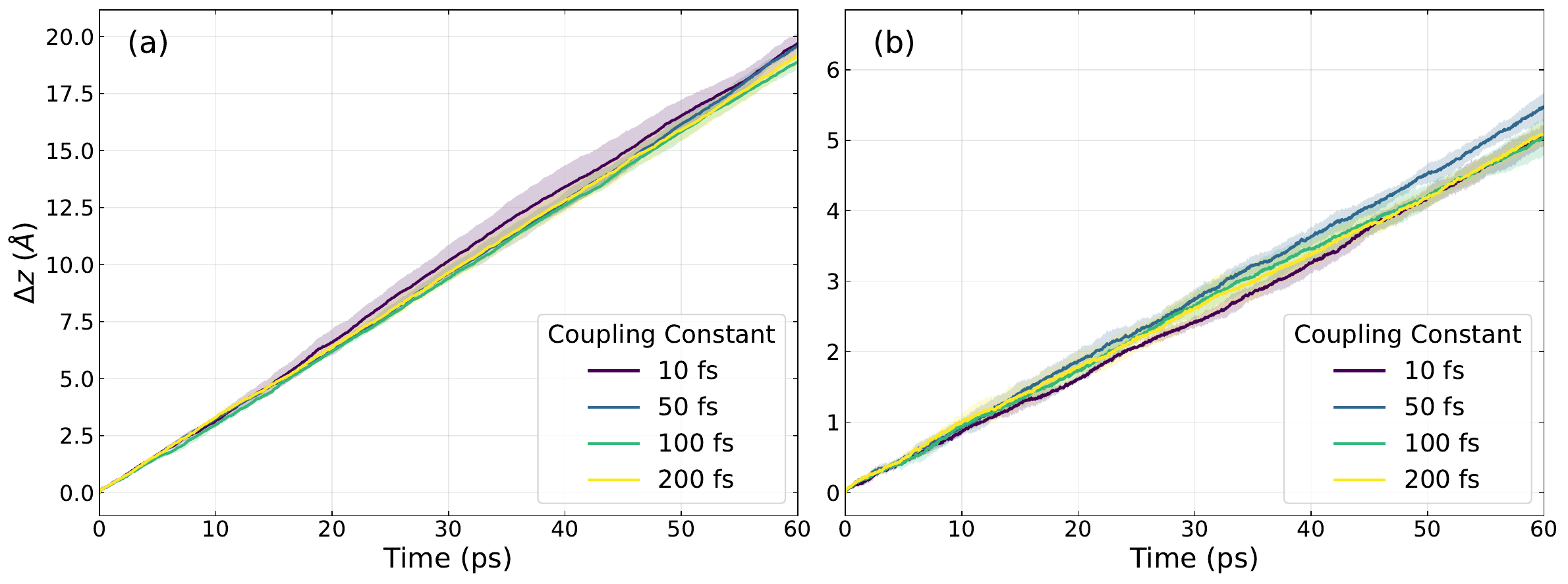}
    \caption{Average displacement of Li ions along $z$ under $E = 0.07$~V/\AA{} at 900~K for thermostat coupling constants of 10, 50, 100, and 200~fs in (a)~amorphous $\text{Li}_{384}\text{P}_{128}\text{O}_{512}$ and (b)~crystalline $\text{Li}_{352}\text{P}_{128}\text{O}_{512}$. Solid lines and shaded regions give the mean and standard deviation of three independent runs.}
    \label{fig:sm-thermostat}
\end{figure}

Standard thermostats are designed for equilibrium conditions. They cannot distinguish the field-induced drift velocity from the random thermal velocity and may therefore suppress both. The microcanonical (NVE) ensemble is the most rigorous choice, but continuous Joule heating restricts it to short runs and weak fields~\cite{nve}. Canonical (NVT) simulations remain usable under stronger fields provided the thermostat error stays small enough not to alter the transport physics~\cite{nvt}. The present study therefore employs NVT MD and validates the thermostat influence explicitly. Simulations under $E = 0.07$~V/\AA{} at 900~K were repeated with Nos\'e-Hoover coupling constants of 10, 50, 100, and 200~fs [Fig.~\ref{fig:sm-thermostat}]. The average Li displacement is consistent across all coupling constants in both phases. The thermostat therefore does not suppress the drift. This is expected because the drift velocity stays below 2\% of the thermal velocity at this temperature.

\section{Zero-field transport}
\label{sec:sm-zerofield}

\begin{figure}[!htb]
    \centering
    \includegraphics[width=0.55\linewidth]{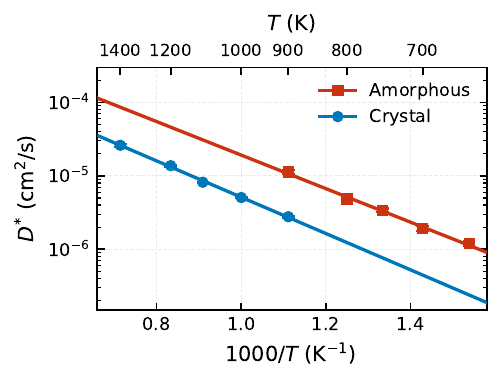}
    \caption{Arrhenius plots of the zero-field tracer diffusion coefficients. Symbols are means over three independent runs and error bars, smaller than the symbols, give the standard error of the mean. Lines are the Arrhenius fits.}
    \label{fig:sm-arrhenius}
\end{figure}

Zero-field MD simulations were performed between 650 and 900~K for the amorphous phase and between 900 and 1400~K for the crystal, with three independent runs per temperature. Each run was equilibrated until the potential energy stabilized, the ballistic regime was excluded~\cite{ballistic}, and the production run was continued until the mean square displacement reached at least 40~\AA$^2$. The equilibrated structures also serve as the starting points of the field runs in the main text. Higher amorphous temperatures were excluded because the diffusivity begins to deviate from Arrhenius behavior there. In both phases the tracer diffusivity $D^{*}$ follows Arrhenius behavior [Fig.~\ref{fig:sm-arrhenius}],
\begin{equation}
  D^{*} = D^{*}_{0} \exp\!\left(-\frac{\Delta H}{k_B T}\right),
\end{equation}
with $\Delta H_{\mathrm{am}} = 0.453 \pm 0.018$~eV and $\Delta H_{\mathrm{cry}} = 0.491 \pm 0.009$~eV. The amorphous activation energy lies about 0.1~eV below typical experimental and computational reports~\cite{kuwata,li,comp_Ea}. We attribute this to the quench rate and to the temperature range analyzed.

\section{Field-direction dependence in the crystal}
\label{sec:sm-directional}

\begin{figure}[htbp]
    \centering
    \includegraphics[width=0.75\linewidth]{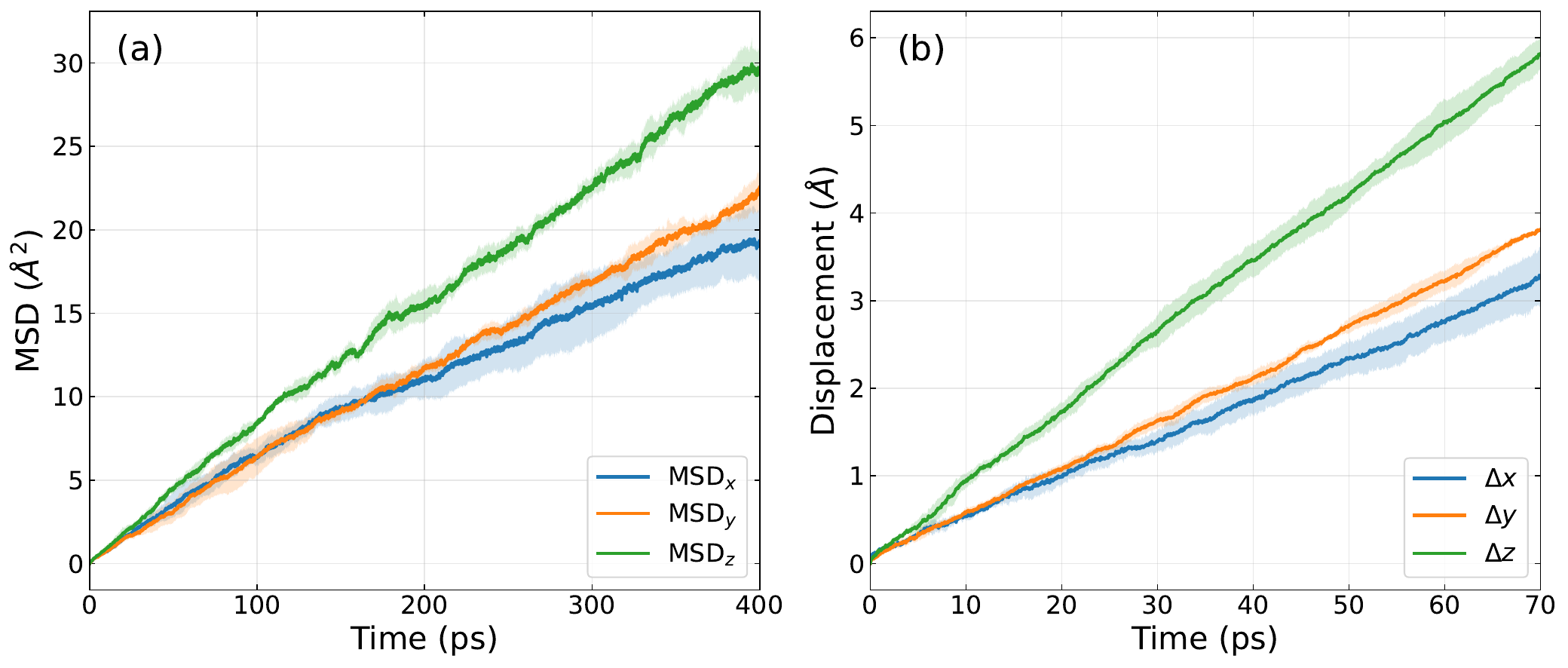}
    \caption{Directional transport in crystalline $\text{Li}_{352}\text{P}_{128}\text{O}_{512}$ at 900~K along $x$, $y$, and $z$ for (a)~the equilibrium mean square displacement and (b)~the average displacement at $E = 0.07$~V/\AA. Solid lines and shaded regions give the mean and standard deviation of three independent runs.}
    \label{fig:sm-directional}
\end{figure}

The field is applied along $z$, which corresponds to the crystallographic $c$ axis ($x \parallel a$, $y \parallel b$) and is the fastest diffusion direction of the anisotropic crystal [Fig.~\ref{fig:sm-directional}]. For the crystal the $x$ and $y$ directions were additionally evaluated at $E = 0.0075$ and $0.07$~V/\AA. At $E = 0.0075$~V/\AA{} the crystal mobilities are, in units of $10^{-4}$~cm$^2$/(V\,s), $\mu_x = 0.44 \pm 0.02$, $\mu_y = 0.65 \pm 0.04$, and $\mu_z = 0.87 \pm 0.04$. Their average, $\mu_{0,\text{cry}} = (0.65 \pm 0.02) \times 10^{-4}$~cm$^2$/(V\,s), is the value quoted in the main text.

\section{Site-occupancy analyses}
\label{sec:sm-occupancy}

In disordered conductors, a nonlinear mobility can also arise if the field redistributes the carriers among energetically inequivalent sites~\cite{heuer2014,wettstein2025}. At zero field the site occupations follow the Boltzmann distribution over the site energies. Under a bias the steady state deviates from thermal equilibrium and the site populations can shift, which by itself can produce a strongly nonlinear mobility~\cite{heuer2014}. An amorphous network, with its broad distribution of site energies, is the natural candidate for this mechanism, which is why we test for it explicitly. All tests are based on the time-averaged Li number density $\rho(\mathbf{r}, E)$, which was accumulated for each field on a regular real-space grid with a spacing of approximately 0.4~\AA{} by binning the Li positions of all production frames of the three independent runs, and smoothed with a periodic Gaussian of width 1~\AA.

\begin{figure}[htbp]
    \centering
    \includegraphics[width=0.42\linewidth]{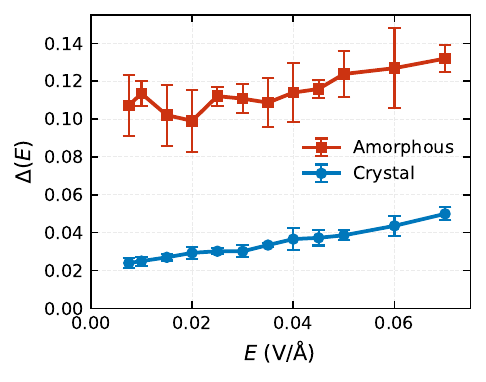}
    \caption{Integrated absolute change of the Li density with respect to zero field as a function of field strength. Means over three replicas, error bars give the standard deviation.}
    \label{fig:sm-delta}
\end{figure}

A first, model-free measure is the integrated absolute density change $\Delta(E) = \int |\rho(\mathbf{r},E) - \rho(\mathbf{r},0)|\, d^3r$, shown in Fig.~\ref{fig:sm-delta}. This measure collects every local change of the density. It therefore responds to a transfer of carriers between sites, but equally to a deformation of the density within each site and to sampling noise. In the crystal, $\Delta$ rises smoothly with the field. The amorphous phase behaves similarly, rising smoothly on top of a much noisier baseline. The baseline is sampling noise, larger in the amorphous phase because the soft wells spread the density over larger volumes. While $\Delta$ alone does not strictly distinguish a deformation of the density within the sites from a transfer between sites, its field-induced increase is small and of the same magnitude in both phases. This already indicates that no strong redistribution takes place in the amorphous network, and the site-resolved tests below confirm it directly.

\begin{figure}[htbp]
    \centering
    \includegraphics[width=\linewidth]{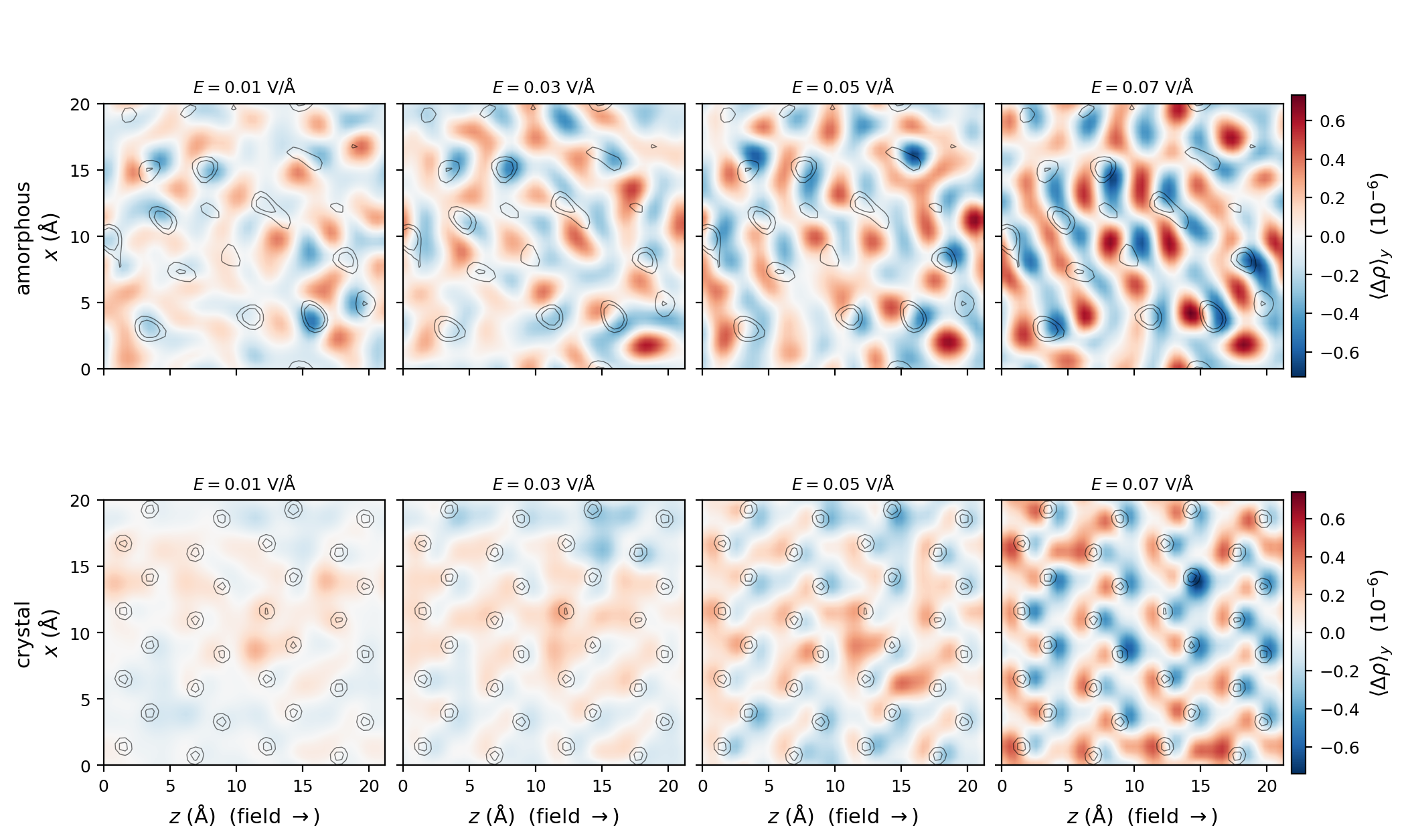}
    \caption{Field-driven change of the Li density, $\langle \Delta\rho \rangle_y = \langle \rho(E) - \rho(0) \rangle_y$, averaged over $y$ and over the three replicas, in the $x$--$z$ plane with the field along $z$, for the amorphous phase (top) and the crystal (bottom). The black lines show two contour levels of the $y$-projected zero-field density, at its 90th and 97th percentile values, so each strong maximum is outlined by an outer and an inner ring. They outline only the strongest projected maxima rather than all sites. In the crystal they correspond to columns of Li sites aligned along $y$, whereas in the amorphous phase sites at different $y$ superpose in the projection.}
    \label{fig:sm-maps}
\end{figure}

Figure~\ref{fig:sm-maps} resolves this growth in space, as a projection of the density change along $y$. In the crystal the Li sites form columns along $y$, so the projection sums them coherently, and gain-and-loss lobes develop symmetrically around each site column with increasing field. This pattern is the direct signature of intra-well polarization, whereas a transfer of carriers between sites would change the amount of density inside the contours. In the amorphous phase the sites sit at random $y$, so the projection superposes unrelated sites and the contours cannot be read as individual sites. At low fields the amorphous panels show no coherent structure beyond the replica sampling noise. At the highest fields, structure correlated with the projected maxima appears, consistent with the stronger intra-well polarization expected for the softer amorphous wells, although the projection precludes a per-site reading. The site-resolved tests below provide that reading.

We note that in Refs.~\cite{heuer2014,wettstein2025} the redistribution mechanism is established for hopping models on disordered lattices and inferred indirectly from experiments, whereas its quantitative contribution to the enhancement of a real amorphous conductor has remained open, because the site energies and occupations of a complex amorphous structure are not directly accessible. The density-based analyses of this section provide such a direct, material-specific test.

\begin{figure}[htbp]
    \centering
    \includegraphics[width=0.9\linewidth]{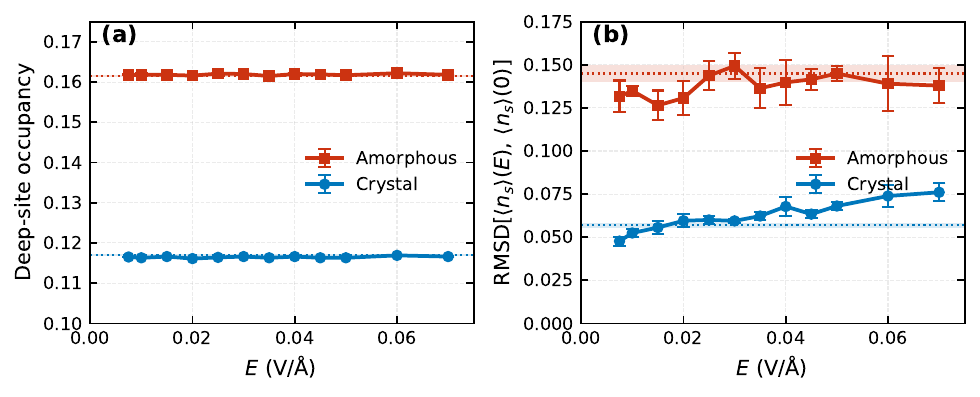}
    \caption{Site-occupancy null tests at 900~K. (a)~Total occupancy of the deep sites, defined as the 10\% of sites with the highest zero-field occupancy, as a function of field. (b)~Root-mean-square change of the per-atom site occupancies $\langle n_s \rangle$ between field and zero field. Error bars give the standard deviation over the three replicas. Dotted lines and shaded bands give the zero-field null, obtained by comparing independent zero-field runs.}
    \label{fig:sm-occ}
\end{figure}

Resolving the change per site requires a site definition that does not rely on crystallographic positions. Sites were therefore defined from the zero-field density itself. Every point in space is assigned to the local maximum of the smoothed density that is reached by always moving uphill, and the region belonging to one maximum is one site. In the crystal this construction recovers exactly the 384 crystallographic Li sites, of which 352 are occupied and 32 are vacant. In the amorphous cell it yields 361 sites for the 384 Li ions, with the slight deficit arising because the smoothing merges a few closely spaced maxima. In the following, every field result is compared with a zero-field null. The null is the difference between two independent zero-field runs, which differ only by sampling noise, so a field effect is significant only if it exceeds this level. The site-integrated occupancies are unchanged from their zero-field values at all fields up to 0.07~V/\AA, the effective number of occupied sites stays constant, and the result is robust to the smoothing width. Particular attention goes to the deep sites, defined as the 10\% of sites with the highest zero-field occupancy. In equilibrium the most occupied sites are the lowest in energy, so these are precisely the traps that the redistribution mechanism predicts to drain under field. Their total occupancy stays on the zero-field null at every field [Fig.~\ref{fig:sm-occ}(a)], with a detection limit of 0.3\%.

Finally, the same conclusion follows from counting atoms directly instead of integrating densities. Each Li ion is assigned in every frame to one of the sites defined above, kept frozen at their zero-field shapes, so the occupancy is a literal atom count. The root-mean-square change of the mean occupancies between field and zero field stays inside the zero-field replica band at every field in the amorphous phase [Fig.~\ref{fig:sm-occ}(b)]. The apparent rise of the crystal curve reflects the shorter production lengths of the high-field runs. When the zero-field reference is truncated to the matched length, the change lies on or below the replica null at every field. The sensitivity of the metric is demonstrated by the crystal curve itself, which responds to this mere change of statistics, and a genuine occupancy shift would add to the replica noise in quadrature and lift the amorphous curve out of its band. Any residual occupancy change below this bound would in any case be far too small to matter. In the redistribution mechanism the enhancement is controlled by the depopulation of the slow, deep sites~\cite{heuer2014}, so a relevant contribution to the measured enhancement would require their population to drop by tens of percent, orders of magnitude above the bound. The site populations are therefore pinned in both phases, and the nonlinear mobility must originate in the hopping kinetics over the field-modified barriers rather than in a redistribution of the carriers.

\FloatBarrier
\section{Hop sampling}
\label{sec:sm-sampling}

\textit{Hop detection.} Hops were detected in the zero-field NVT MD at 900~K. Most of the time a Li rattles in its cage, and only occasionally does it jump to a new position. To separate the two, we compare at every time $t$ the mean position of each Li over the next 0.3~ps with its mean position over the previous 0.3~ps. The distance between the two, $\delta(t)$, stays small while the Li rattles in one cage and rises when it moves to another. A jump is a period in which $\delta$ exceeds 1~\AA. Two jumps of the same Li less than 0.75~ps apart are merged into one hop. This turns a quick jump forth and back into a single short excursion, and it joins two jumps through a short stop between sites into one hop. The hop time $t_{J}$ is the moment of largest $\delta$ within the hop. The hop length is the distance between the mean positions of the Li during its stays before and after the hop, where a stay is the time between two consecutive hops.

\textit{Choice of the parameters.} The averaging window must be long enough to average out the rattling in a cage. In the amorphous phase the rattling ends at the caging time of 0.18~ps, where the local slope of the Li mean-square displacement has its minimum, and the window of 0.3~ps is about twice this time. The merging time was set in the crystal, where the sites are known. There a Li stays between two jumps either at a lattice site, within 0.7~\AA{} of it, or at a stop between sites. Stays at stops are short and stays at sites are long, and the two distributions of stay durations cross at 0.75~ps. A stay shorter than 0.75~ps is therefore most likely a stop between sites, and the jumps before and after it form one hop between two sites. The same merging time is used in both phases.

\textit{Selection.} In the amorphous phase, the hops were restricted to MD hop lengths of 1.5--3.5~\AA. In each 0.5~\AA{} interval, hops were drawn at random and attempted in draw order until 25 were accepted, which gives 100 hops. In the crystal, hops were drawn at random from all detected events until 50 were accepted. The outcome of every attempted hop was recorded.

\textit{Endpoints and relaxation.} The initial and final site centers of a hop are the mean positions of the Li over the 0.3~ps before and after $t_{J}$. The initial frame is the last MD frame before $t_{J}$ in which the Li is at or behind the initial site center along the hop direction, and the final frame the first frame after $t_{J}$ in which it is at or beyond the final site center. The initial frame is relaxed. The final endpoint is built from the relaxed initial structure by moving only the hopping Li to its position in the relaxed final frame, and it is relaxed again. Both relaxations converge the forces to 0.005~eV/\AA.

\textit{NEB and acceptance.} The minimum energy path is obtained with climbing-image NEB with 7 intermediate images, converged to 0.01~eV/\AA, and $\Delta\mathbf{P}^{S}$ is integrated along the band from the Born effective charges of every image. If the band passes through a state lower than an endpoint, that state is relaxed and becomes the new endpoint, and the NEB is repeated on the part of the band that contains the hop. A hop is rejected for one of the following reasons.
\begin{itemize}
\item \textit{Collapse.} After the relaxation the Li moves less than half the distance between its positions in the relaxed initial and final MD frames, so the two endpoints are not distinct minima. The destination may be a shallow site that does not survive the relaxation, or the hop may belong to a reorganization of the environment in which the destination exists only in the moving environment.
\item \textit{Destination blocked.} Another Li lies within 1.2~\AA{} of the destination in the relaxed initial structure, and the final endpoint built by moving only the hopping Li would put two Li on top of each other. This happens when another Li must move for the hop to take place, or when the relaxation moves another Li into the destination. The value is half of 2.4~\AA, the shortest Li--Li distance found in more than 5\% of the nearest-neighbor pairs of the relaxed amorphous structures.
\item \textit{Barrier below $k_{B}T$.} The forward or backward barrier is below 78~meV. Such an endpoint is a minimum so shallow that the Li does not stay in it at 900~K, so the hop is not a jump between two sites in which the Li resides, the process studied here.
\item \textit{Detected Li not the largest mover.} Another atom moves more than the detected Li in every part of the band, so the band describes a different event. Other atoms, including other Li, may still move in an accepted band.
\item \textit{Not converged.} The NEB band did not converge, because the endpoints also differ in the positions of other atoms in a way that a single band between them cannot represent.
\end{itemize}

\begin{table}[htbp]
\caption{Outcome of the attempted hops as percentages of the attempts, up to the last accepted hop of each ensemble. The amorphous attempts are also given by MD hop length. The last row gives the number of attempted hops.}
\label{tab:sm-outcome}
\begin{ruledtabular}
\begin{tabular}{lrrrrrr}
 & \multicolumn{4}{c}{Amorphous, MD hop length (\AA)} & Amorphous & Crystal \\
 & 1.5--2.0 & 2.0--2.5 & 2.5--3.0 & 3.0--3.5 & total & \\ \hline
Accepted & 6\% & 10\% & 12\% & 15\% & 10\% & 77\% \\
Collapse & 72\% & 60\% & 40\% & 24\% & 55\% & 3\% \\
Destination blocked & 6\% & 19\% & 33\% & 49\% & 22\% & 15\% \\
Barrier below $k_{B}T$ & 14\% & 10\% & 11\% & 8\% & 11\% & 5\% \\
Not the largest mover & 1\% & 1\% & 1\% & 1\% & 1\% & 0\% \\
Not converged & 0\% & 1\% & 2\% & 4\% & 1\% & 0\% \\
\hline
Attempted hops & 392 & 251 & 210 & 170 & 1023 & 65 \\
\end{tabular}
\end{ruledtabular}
\end{table}

\textit{Outcome.} Table~\ref{tab:sm-outcome} lists the outcome of all attempted hops. In the crystal, 77\% of the hops are accepted. Apart from the ten blocked hops discussed below, the failures are five hops with an endpoint between the lattice sites, either in the MD frame or after the relaxation, which collapse or have a barrier below $k_{B}T$. In the amorphous phase, only 10\% of the hops are accepted, and the main reason for failure depends on the MD hop length. Short hops mostly collapse (72\% at 1.5--2.0~\AA{} against 24\% at 3.0--3.5~\AA), whereas long hops are more often blocked (49\% at 3.0--3.5~\AA{} against 6\% at 1.5--2.0~\AA). No accepted band in either phase contains an intermediate minimum deeper than $k_{B}T$.

In a blocked hop, another Li still occupies the destination before the hop. In many cases this Li moves away at about the same time as the hopping Li arrives. In the amorphous MD, the blocking Li itself hops within 0.3~ps before or after the hop in 58\% of the cases. For comparison, an arbitrary Li hops within a given 0.6~ps window with a probability of only 8\%, as estimated from the average hop rate of a Li. The hop of the blocking Li is thus correlated with that of the hopping Li.

In the crystal, where the lattice sites allow the motion to be followed directly, the blocking Li leaves the destination for a neighboring vacancy before the hopping Li arrives in all ten blocked hops. In four of them, it is already on its way out in the initial frame, and the relaxation pulls it back into the destination. The two Li could still move in a concerted hop, in which several ions migrate simultaneously~\cite{he2017}. To test this, we computed an NEB in which the hopping Li and the Li in its way move together to their positions in the final MD frame. In seven of the ten blocked hops, this band separates into consecutive steps with clear barriers of their own, so the Li move one after the other. In the other three, the band has a single saddle. For these we also computed the path in which the Li move one after the other, with one NEB from the initial state to an intermediate state in which only the blocking Li has moved and a second NEB from this intermediate state to the final state. Its highest barriers, 391, 548, and 369~meV, are lower than those of the two-ion bands, 486, 740, and 514~meV, so a concerted hop is possible there but less likely. Most blocked crystal hops are therefore consecutive single-ion hops of the kind already in the ensemble. Even if the three single-saddle hops were counted as concerted hops with the same weight as the accepted ones, $\langle Q^{\mathrm{eqv}}\rangle$ of the crystal would rise only from $+1.10$ to $+1.14$ and the crystal model at 0.07~V/\AA{} from 1.33 to 1.35.

The accepted ensemble therefore represents hops with two stable endpoints that can be produced by moving only the hopping Li. It under-represents hops that require other Li to move, which are more common among long hops. In the crystal, counting such hops as concerted would raise $\langle Q^{\mathrm{eqv}}\rangle$ slightly, but this has not been established for the amorphous phase. Correlated or dynamically stabilized events are thus not fully represented in the hopping model. The broad environment contribution, however, is already present within the accepted hops.

\section{Hop-length dependence}
\label{sec:sm-hopdist}

\begin{figure}[H]
    \centering
    \includegraphics[width=0.55\linewidth]{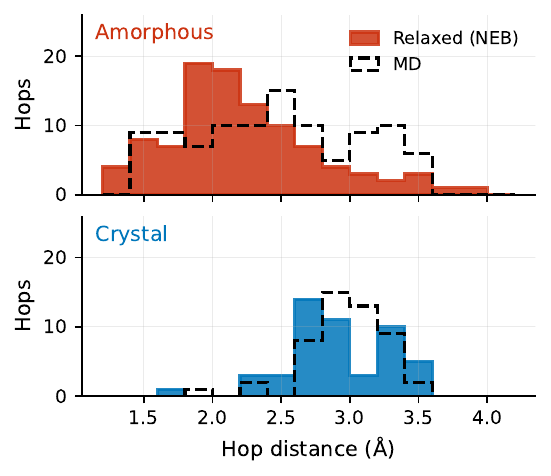}
    \caption{Distribution of the relaxed hop distances of the NEB ensembles, 100 amorphous and 50 crystal hops (filled), and of their hop lengths detected in the MD (dashed).}
    \label{fig:sm-hopdist}
\end{figure}

Figure~\ref{fig:sm-hopdist} shows the hop distances of the NEB ensembles, measured as the displacement of the migrating atom between the two endpoints of its accepted band, together with the hop lengths detected in the MD. The amorphous hops are shorter, with a median of 2.1~\AA{} against 2.9~\AA{} in the crystal. In the amorphous phase the MD hop length is only partly preserved by the relaxation. Most hops relax to 1.9--2.6~\AA, because the MD length includes thermal displacement and can span merged moves, and because the environment reorganizes during the relaxation. The hops were drawn evenly over MD hop lengths of 1.5--3.5~\AA, whereas most hops detected in the MD are short. Of the 10\,287 detected hops of 1.5--3.5~\AA, 4\,414 (42.9\%), 3\,153 (30.6\%), 1\,915 (18.6\%), and 805 (7.8\%) fall in the four intervals. In the hopping model of the End Matter, each amorphous hop is therefore weighted by the fraction $w_{i}$ of its interval, so that the ensemble follows the hop-length distribution of the MD. A hop keeps the weight of its MD interval whatever its relaxed length, because the hops were drawn by MD hop length.

Table~\ref{tab:sm-length} gives the statistics of the amorphous ensemble and the hopping model of the End Matter for each quarter of relaxed hop length. In every quarter the spread of $Q^{\mathrm{eqv}}$ is large, and the model with the full polarization change gives a much larger enhancement than with the migrating ion alone. The enhancement is by far the largest in the longest quarter, because the jumps are longer, the environment work spreads most, and, in a sample of 25 hops, a few strongly assisted hops can dominate. The conclusion of the main text, that the environment work rather than the migrating ion produces the stronger amorphous response, thus holds in every range of hop length.

\begin{table}[H]
\caption{Amorphous ensemble by quarter of relaxed hop length (25 hops each). Uncertainties are bootstrap standard errors, SD is the hop-to-hop standard deviation, and $e\,Q_{\mathrm{env}}|\mathbf{d}^{S}_{\mathrm{MA}}|$ is the environment work per unit field along the hop in $e$\,\AA. The last two columns give $\mu/\mu_{0}$ at 0.07~V/\AA{} of the hopping model for the hops of each quarter on their own, with the full polarization change and with the migrating ion alone (MD: 1.69). These columns use equal weights.}
\label{tab:sm-length}
\begin{ruledtabular}
\begin{tabular}{lccccccc}
Hop length (\AA) & Hops & $\langle Q^{\mathrm{eqv}}\rangle$ & SD of $Q^{\mathrm{eqv}}$ & $\langle Q_{\mathrm{env}}\rangle$ & SD of $e\,Q_{\mathrm{env}}|\mathbf{d}^{S}_{\mathrm{MA}}|$ & Model, full & Model, ion only \\ \hline
1.22--1.87 & 25 & $+1.43 \pm 0.13$ & 0.90 & $+0.42 \pm 0.12$ & 0.51 & 1.40 & 1.08 \\
1.87--2.12 & 25 & $+1.32 \pm 0.22$ & 1.28 & $+0.30 \pm 0.21$ & 1.07 & 1.64 & 1.12 \\
2.15--2.53 & 25 & $+0.88 \pm 0.11$ & 0.67 & $-0.12 \pm 0.11$ & 0.73 & 1.45 & 1.21 \\
2.53--3.81 & 25 & $+1.01 \pm 0.23$ & 1.34 & $+0.04 \pm 0.23$ & 2.03 & 3.23 & 1.35 \\
\end{tabular}
\end{ruledtabular}
\end{table}

\section{Environment motion during a hop}
\label{sec:sm-envdisp}

\begin{figure}[htbp]
    \centering
    \includegraphics[width=0.7\linewidth]{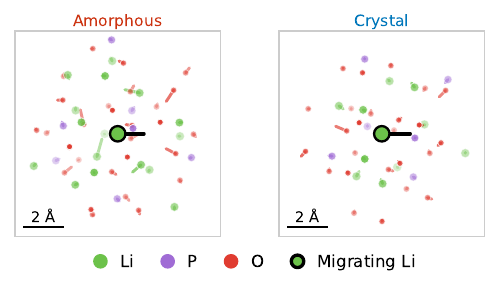}
    \caption{Displacements from the initial state to the saddle point for the amorphous and crystal hops of the text, in the plane of the hop and the field axis $z$, rotated so that the hop points to the right. The black line is the path of the migrating ion. Colored lines are the displacements of the environment atoms within 5~\AA{} amplified threefold for visibility and faded with depth.}
    \label{fig:sm-hopmap}
\end{figure}

The collective motion behind the environment term is quantified by the summed displacement of all environment atoms between the initial state and the saddle point, $\sum_{a \neq \mathrm{MA}} |\Delta\mathbf{r}^{S}_{a}|$. Averaged over the NEB ensembles, it amounts to 26~\AA{} per hop in the amorphous phase against 8~\AA{} in the crystal, the values quoted in the main text. Figure~\ref{fig:sm-hopmap} illustrates this contrast for an amorphous hop and a crystal hop with typical environment motion (21 and 9~\AA) and similar saddle-point displacements (1.38 and 1.49~\AA). The amorphous environment visibly follows the migrating ion, whereas the crystal environment barely moves.

The rigid-unit test of the main text separates the two ways in which this motion polarizes the environment, deformation or rotation of the $\text{PO}_4$ units and relative displacement of Li ions and $\text{PO}_4$ units. Each O atom is assigned to its nearest P atom in the initial state, and at every NEB image all atoms of a unit are given the displacement of the unit's center of mass. The polarization change of Eq.~(3) of the main text is then integrated along the band with these displacements and the full BEC tensors, while the Li ions of the environment keep their actual displacements. Each unit thus responds as a whole with the summed BEC of its atoms, and neither its deformation nor its rotation enters. The difference from the actual $Q_{\mathrm{env}}$ is the contribution of motion within the units.

\section{Validation of the equivalent charges}
\label{sec:sm-robust}

\begin{figure}[H]
    \centering
    \includegraphics[width=0.55\linewidth]{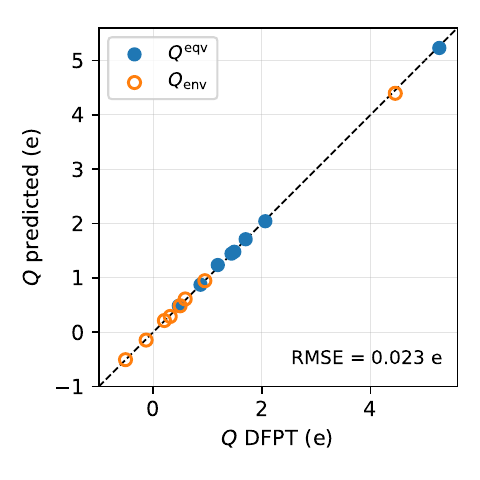}
    \caption{Path-integrated $Q^{\mathrm{eqv}}$ (filled) and $Q_{\mathrm{env}}$ (open) predicted by Equivar against DFPT for eight path directions in the independent 125-atom amorphous cell. The dashed line marks perfect agreement.}
    \label{fig:sm-a125}
\end{figure}

The path-integrated charges of the main text were validated directly against DFPT on the independent 125-atom amorphous structure introduced in the model-validation section above. Four NEB paths were computed with GRACE, chosen such that $Q_{\mathrm{env}}$ covers negative and positive values, from $-0.5$ to $+4.5$. The BEC tensors of all images were recomputed with DFPT, and the polarization change was integrated once with the DFPT tensors and once with Equivar on the same geometries, so the comparison isolates the charge model. The BEC components agree with an RMSE of $0.023\,e$. For the eight path directions evaluated, $Q^{\mathrm{eqv}}$ agrees with an RMSE of $0.023\,e$ and $Q_{\mathrm{env}}$ with $0.025\,e$ [Fig.~\ref{fig:sm-a125}]. The environment contribution is therefore not a model artifact.

\section{Orientation average in the hopping model}
\label{sec:sm-orient}

Equation~(A3) of the End Matter follows from Eq.~(A2) by averaging each hop over all orientations. Hop $i$ contributes $d_{i}\cos\theta_{i}\sinh(a_{i}\cos\phi_{i})$ with $a_{i} = \Omega\,|\Delta\mathbf{P}_{i}|\,E/k_{B}T$, where $\theta_{i}$ is the angle between the hop and the field and $\phi_{i}$ the angle between $\Delta\mathbf{P}_{i}$ and the field. The angle $\alpha_{i}$ between the hop and $\Delta\mathbf{P}_{i}$ is fixed by the hop. Rotating the hop as a whole over all orientations is equivalent to keeping the hop fixed and letting the field point in all directions with equal probability. We take the polar axis along $\Delta\mathbf{P}_{i}$ and the hop direction in the $xz$ plane, so that the field direction has the polar angle $\phi$ and the azimuth $\psi$, and
\begin{equation}
  \cos\theta = \cos\alpha_{i}\cos\phi + \sin\alpha_{i}\sin\phi\cos\psi .
\end{equation}
Averaging over the solid angle $\sin\phi\,d\phi\,d\psi$,
\begin{equation}
  \langle v_{i}\rangle \propto \frac{d_{i}}{4\pi}\int_{0}^{2\pi}\!\! d\psi \int_{0}^{\pi}\!\! \sin\phi\, d\phi\, \left(\cos\alpha_{i}\cos\phi + \sin\alpha_{i}\sin\phi\cos\psi\right) \sinh(a_{i}\cos\phi).
\end{equation}
The work does not depend on $\psi$, so the second term vanishes with $\int_{0}^{2\pi}\cos\psi\, d\psi = 0$. The part of the hop perpendicular to $\Delta\mathbf{P}_{i}$ moves the ion forward and backward along the field equally often for a given work and does not contribute. With $x = \cos\phi$, the first term gives
\begin{equation}
  \langle v_{i}\rangle \propto \frac{d_{i}\cos\alpha_{i}}{2} \int_{-1}^{1} x \sinh(a_{i} x)\, dx = d_{i}\cos\alpha_{i}\left(\frac{\cosh a_{i}}{a_{i}} - \frac{\sinh a_{i}}{a_{i}^{2}}\right),
\end{equation}
and summing over the hops gives Eq.~(A3). A direct numerical average of Eq.~(A2) over the field directions gives the same curve.

\bibliography{references}